\documentclass[10pt]{iopart}
\usepackage{iopams, graphicx,url, bm, color, tikz, pgfplots, hyperref, times}
\usepackage{siunitx}[=v2]
\expandafter\let\csname equation*\endcsname\relax
\expandafter\let\csname endequation*\endcsname\relax
\usepackage{amsmath, amsfonts, amssymb}
\usepackage[switch]{lineno}
\usepackage[style=numeric-comp, backend=biber, sorting=none, bibencoding=utf8, url=false, doi=true, eprint=false, isbn=false, firstinits, maxcitenames=2, maxbibnames=3]{biblatex}

\let\originalleft\left
\let\originalright\right
\renewcommand{\left}{\mathopen{}\mathclose\bgroup\originalleft}
\renewcommand{\right}{\aftergroup\egroup\originalright}

\hypersetup{colorlinks=true, linkcolor=blue, citecolor=blue, urlcolor=blue}

\providecommand{\e}{\ensuremath{\textrm{e}}}
\providecommand{\Tg}{\ensuremath{T_{\textrm{g}}}}
\providecommand{\Ntwo}{\ensuremath{\textrm{N}_2}}
\providecommand{\Otwo}{\ensuremath{\textrm{O}_2}}
\providecommand{\NtwoP}{\ensuremath{\textrm{N}_2^+}}
\providecommand{\OtwoP}{\ensuremath{\textrm{O}_2^+}}
\providecommand{\OtwoM}{\ensuremath{\textrm{O}_2^-}}
\providecommand{\OM}{\ensuremath{\textrm{O}^-}}
\providecommand{\Oone}{\ensuremath{\textrm{O}}}
\providecommand{\NtwoO}{\ensuremath{\textrm{N}_2\textrm{O}}}
\providecommand{\Etd}{\ensuremath{\textrm{E}^\prime}}
\providecommand{\M}{\ensuremath{\textrm{M}}}
\providecommand{\OthreeM}{\ensuremath{\textrm{O}_3^-}}
\providecommand{\NfourP}{\ensuremath{\textrm{N}_4^+}}
\providecommand{\OfourP}{\ensuremath{\textrm{O}_4^+}}
\providecommand{\bme}{\ensuremath{\bm{E}}}

\begin{document}

\title[Positive streamer propagation over non-planar dielectrics]{Evolution of positive streamers in air over non-planar dielectrics: Experiments and simulations}

\author{H K H Meyer}
\address{SINTEF Energy Research, Sem S\ae lands vei 11, 7034 Trondheim, Norway.}
\ead{hans.meyer@sintef.no}

\author{R Marskar}
\address{SINTEF Energy Research, Sem S\ae lands vei 11, 7034 Trondheim, Norway.}
\ead{robert.marskar@sintef.no}

\author{F Mauseth}
\address{NTNU - Norwegian University of Science and Technology}
\ead{frank.mauseth@ntnu.no}

\date{\today}

\begin{abstract}
  We study positive streamers in air propagating along polycarbonate dielectric plates with and without small-scale surface profiles.
  The streamer development was documented using light-sensitive high-speed cameras and a photo-multiplier tube, and the experimental results were compared with 2D fluid streamer simulations.
  Two profiles were tested, one with 0.5 mm deep semi-circular corrugations and one with 0.5 mm deep rectangular corrugations.
  A non-profiled surface was used as a reference.
  Both experiments and simulations show that the surface profiles lead to significantly slower surface streamers, and also reduce their length.
  The rectangular-cut profile obstructs the surface streamer more than the semi-circular profile.
  We find quantitative agreement between simulations and experiments.
  For the surface with rectangular grooves, the simulations also reveal a complex propagation mechanism where new positive streamers re-ignite inside the surface profile corrugations.
  The results are of importance for technological applications involving streamers and solid dielectrics.
\end{abstract}

\maketitle
\ioptwocol

\section{Introduction}
Streamer discharges \cite{Nijdam2020} are low-temperature filamentary plasma, driven by electron impact ionization and charge separation in the streamer head.
They can appear when the electric field is locally higher than the breakdown field.
Once streamers have formed, the self-enhanced field at the streamer tips permits them to propagate in background fields lower than the breakdown field.
For example, the breakdown field in atmospheric air is $E_{\textrm{br}} \sim 30\,\si{kV/cm}$, but positive streamers can propagate in fields down to $\sim 5\,\si{kV/cm}$.

Streamers can propagate in bulk gases but also along dielectric surfaces.
Although the governing physics is mostly the same in both cases, there are some differences between the two:
1) Dielectric polarization enhances the field between the streamer and the dielectric, and streamers therefore tend to be \emph{surface-hugging}.
2) Streamers typically propagate faster along a dielectric surface than in the surrounding gas \cite{allen_streamer_1999,allen_surface_2001,sobota_speed_2008,li_computational_2020}.
3) Positive streamers are accompanied by a cathode sheath closest to the dielectric \cite{soloviev_mechanism_2014, Marskar2019a, li_computational_2020,meyer_streamer_2019}.
This sheath is analogous to the cathode sheath on electrodes.
It appears because secondary electrons from ionization and surface emission are low-energy electrons that move away from the surface and thus do not ionize the gas closest to the dielectric.
4) Streamers can be influenced by emission of secondary electrons from the dielectric surface, but in atmospheric air the overall effect on the streamer velocity and range is rather small \cite{meyer_streamer_2019}.
5) Surface charge also affect the streamer behavior, and can for example lead to inception of airborne streamers from the surface \cite{florkowski_imaging_2021}.
However, charging of dielectric surfaces exposed to positive streamers mainly occurs through ion transport in the streamer wake, i.e. from the channel and onto the surface \cite{meyer_streamer_2019}.
Since the ions only move a short distance, surface charging essentially do not affect ionization process in the streamer head \cite{li_computational_2020}.

Streamer propagation along dielectrics is important in several research fields.
Discharge propagation along gas-dielectric interfaces is often critical for medium voltage (MV) switchgear, which usually contain solid insulating components like supports, shafts, and barriers.
Aerodynamic plasma actuators also operate with surface streamers that propagate along the airfoil, usually in the form of a dielectric barrier discharge (DBD) \cite{Boeuf2005}.
Another example is plasma wound healing, in which the dielectric is human tissue \cite{Laroussi2018}.
The shape, orientation, profile, or roughness of a dielectric surface are all important factors for streamer propagation along surfaces.
Positive streamers in air, for example, have radii on the order of $0.1\text{-}1\,\si{mm}$ \cite{Briels2008}, and are thus easily manipulated by modifying the dielectric surface structure on the micron or millimeter scale. 

There are several studies that address streamer propagation over planar surfaces \cite{marskar_adaptive_2019, Soloviev2009, Soloviev2017, li_computational_2020}, but studies of streamer propagation over non-planar surfaces are scarce.
An investigation of positive streamer propagation over a profile with semi-circular corrugations was presented in \cite{meyer_streamer_2020} by the authors.
Comparing both experiments and simulations, we showed that surface profiles reduce both the velocity and range of positive streamers.
Moreover, the authors experimentally compared propagation over surfaces with 0.5 mm deep rectangular cut corrugations with a flat surface in \cite{meyer_streamer_2022}, and found that positive streamer propagation was more restricted over such surfaces than over semi-circular profiled or flat surfaces.
It was also confirmed in \cite{meyer_streamer_2022} that positive streamers follow semi-circular corrugations closely, as predicted by simulations in \cite{meyer_streamer_2020}.
\textcite{wang_effect_2021} numerically examined the profile effect for both convex and concave corrugations.
\textcite{pritchard_streamer_2002} found that streamers propagating along dielectrics with various high voltage insulator shed designs required higher background electric field strengths to sustain propagation than for streamers along plain insulators.

In this paper we perform both computer simulations and high-speed imaging of experiments of positive streamers propagating over various dielectric surface profiles.
We extend the analysis in \cite{meyer_streamer_2020} and \cite{meyer_streamer_2022}, mainly by including simulations and more experiments of the rectangularly profiled surface, by using more elaborate plasma models, and by using a consistent impulse voltage curve in experiments and simulations.
The goal of this article is two-fold.
Firstly, we wish to establish a deeper understanding of positive streamer propagation along various surface profiles.
Secondly, we wish to take steps towards computational validation of the discharge model through an apples-to-apples comparison with experiments.

The outline of this paper is as follows:
In \sref{sec:experiments} we outline the experimental setup and procedure.
We discuss the numerical approach in \sref{sec:computer_simulations}, and present our results in \sref{sec:results}.
A summary of the paper findings is then presented in \sref{sec:conclusions}.

\section{Experimental methods}

\label{sec:experiments}
\subsection{Test objects}
\begin{figure}[h!t!b!]
  \centering
  \includegraphics[width=0.45\textwidth,trim={0cm 0cm 2cm 0cm},clip]{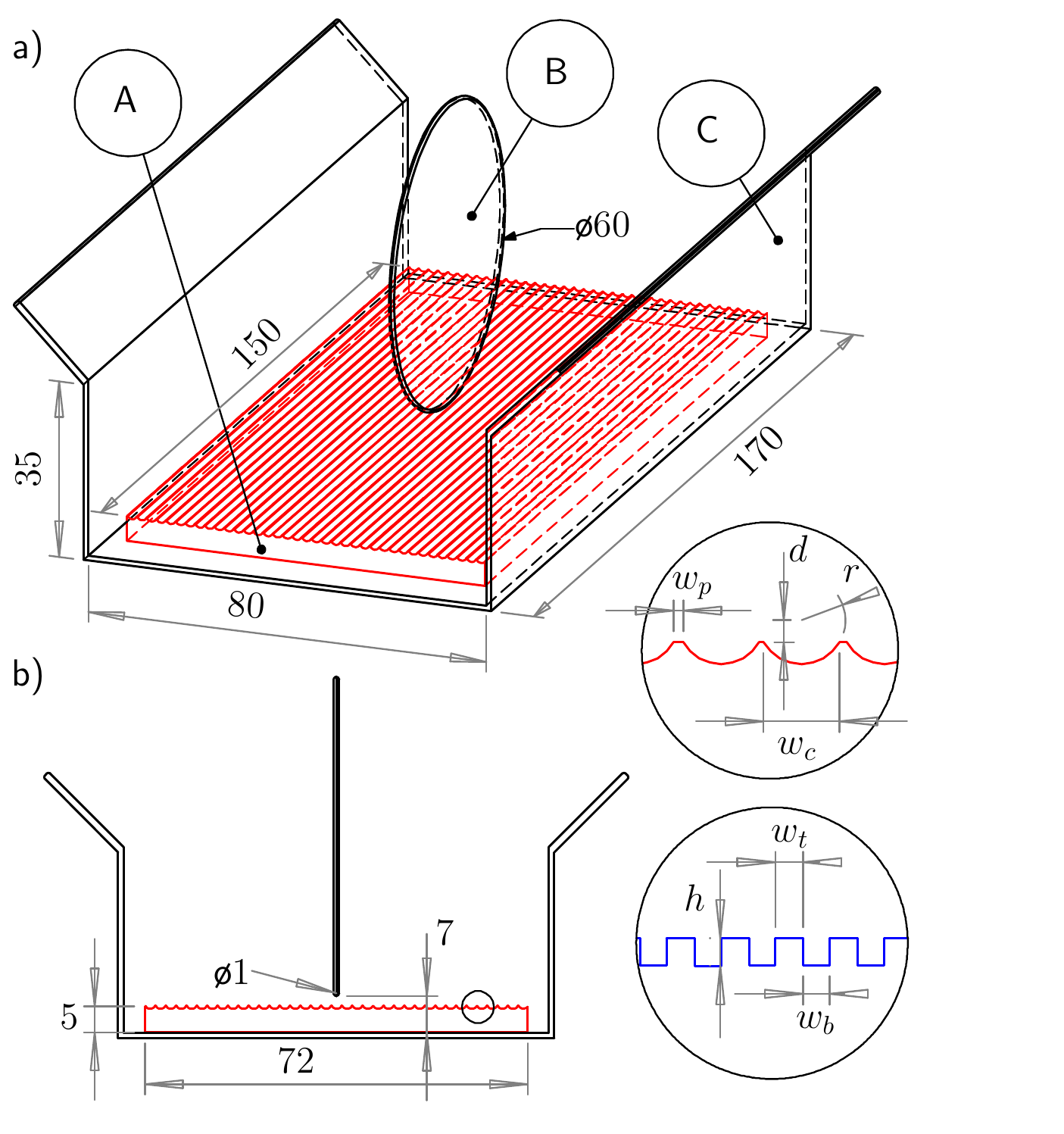}
  \caption{Electrodes and dielectric test objects
    (a) Setup viewed from an angle, showing polycarbonate surface with 0.5 mm semi-circular corrugations outlined in red (A), the grounded aluminium casing (B), and the HV disk electrode (C).
    (b) Front view, with details showing dimensions of the semi-circular surface profile in red and rectangular surface profile in blue. All indicated dimensions are given in millimeters.}
  \label{fig:profiles}
\end{figure}

\Fref{fig:profiles} illustrates the experimental setup, including the dimensions of the electrodes and the surface profile details.
Three polycarbonate (Lexan) plate dielectrics of $5\,\si{mm}\times72\,\si{mm}\times150\,\si{mm}$ and relative permittivity $\epsilon_r = 3$ were used as test objects:

\begin{enumerate}
\item A flat plate.
\item A plate with semi-circular corrugations as shown in \fref{fig:profiles}a) and \fref{fig:profiles}b).
  This profile is a reproduction of the test object in \cite{meyer_streamer_2020}, and has a 20 \% greater surface area than the plain surface.

\item A plate with rectangular corrugations.
  This surface has a 110 \% larger surface area than the flat plate.
  This surface profile is shown in blue color in \fref{fig:profiles}c).
  The rectangular profile restricts streamer propagation along the surface as shown in \cite{meyer_streamer_2022} .
\end{enumerate}

For the profiled plates, the surface profiles consist of $\sim 500\,\mu\si{m}$ deep corrugations, which were drilled using a bore head.
The dimensions of the profiled surface were measured with a Bruker ContourGTK profilometer and averaged, see \tref{tab:dimensions}.

\begin{table}[h!t!b!]
  \caption{
    \label{tab:dimensions}
    Average measured surface dimensions of the profiled test objects (see \fref{fig:profiles}).
    All quantities are given in microns ($\mu\si{m}$).
  }
  \begin{indented}
  \item[]
    \centering
    \begin{tabular}{@{}llll|lll}
      \br
      \multicolumn{4}{c|}{Semi-circular} & \multicolumn{3}{c}{Rectangular} \\
      \mr
      $w_c$ & $w_p$ & $d$  & $r$  & $w_t$ & $w_b$ & $h$  \\
      \mr
      1718  & 227  &  459 & 974  & 524     & 491     & 546    \\
      \br
    \end{tabular}
  \end{indented}
\end{table}

An aluminum casing and a disk-shaped brass electrode were used as ground and high voltage (HV) electrodes.
The disk shape was used as its electrical field distribution resembles the 2D planar simulation field, and it provides a small streamer inception region which facilitates imaging.

\begin{figure}[h!t!b!]
  \centering
  \includegraphics[width=0.45\textwidth,trim={0.7cm 1.2cm 0.5cm 0.8cm}, clip]{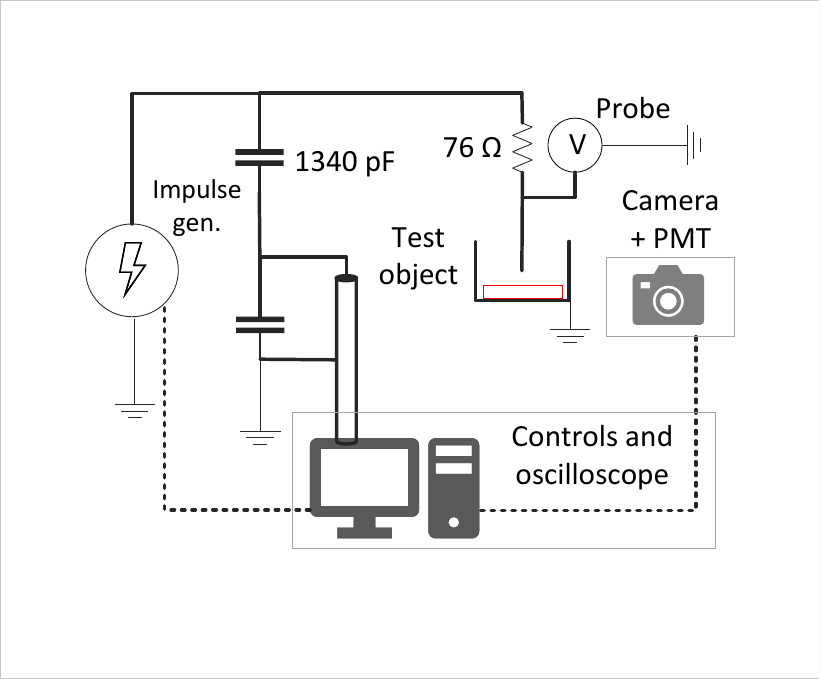}
  \caption{Streamer imaging setup. Voltage impulses were applied with a impulse generator to the test object in \fref{fig:profiles}.
    The dielectric surface is highlighted in red.
    The voltage shape was measured both with a capacitive divider and a high voltage probe.}
  \label{fig:schematic}
\end{figure}

A voltage impulse was applied to the disk electrode using one stage of a 1.2 MV lightning impulse (LI) generator, see \fref{fig:schematic}.
A resistor in series with the test object limited the discharge energy.
The voltage was measured directly on the test object using a North Star PVM-100 high-voltage probe.

\begin{figure}[h!t!b!]
  \centering
  \includegraphics[width=0.45\textwidth]{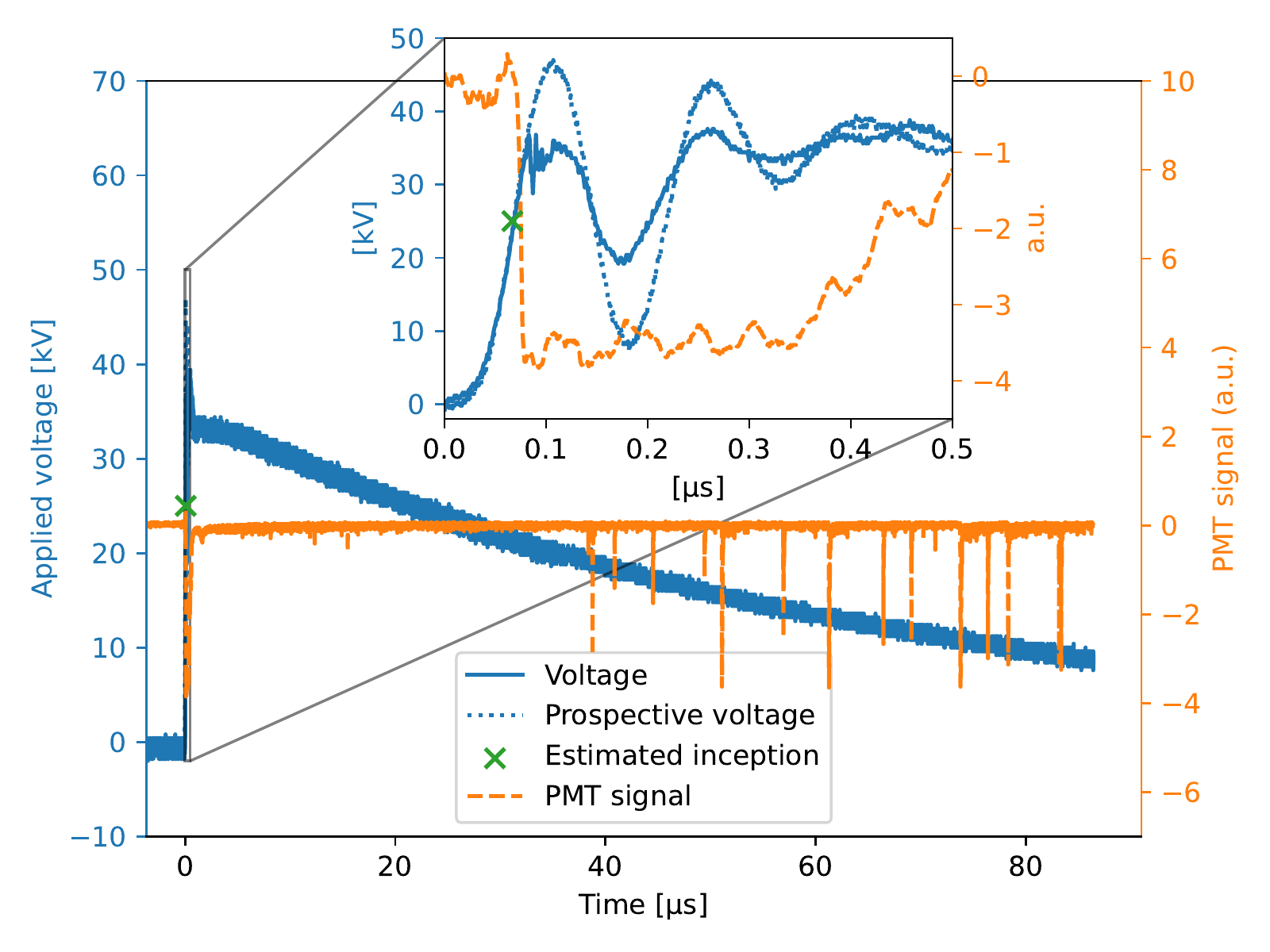}
  \caption{
    Example voltage shape, voltage drop and PMT signals during streamer propagation for the flat dielectric surface.
    The streamer activity was recorded with a \SI{35}{\kilo\volt} charging voltage.
    The inset shows the events around the rising voltage flank.
    The estimated inception time and voltage is also shown.}
  \label{fig:test_object_voltage}
\end{figure}

Since the computer simulations are sensitive to the applied voltage, we used a simulation voltage similar to the prospective voltage curve in \fref{fig:test_object_voltage}.
The prospective voltage curve was measured with the voltage probe using a charging voltage which was too low to cause inception, and then scaled up based on generator charging voltages.
The rise time (10 to 90 \%) of the initial voltage front was $53\,\si{ns}$, while the impulse half-value time was around $50\,\mu\si{s}$.
The short rise time faciliated imaging, but resulted in a voltage overshoot due to oscillations in the circuit, which produced a peak voltage of $48\,\si{kV}$ over the test object for a $35\,\si{kV}$ charging voltage. 
The actual voltage measured over the gap depended on the streamer activity as shown in \fref{fig:test_object_voltage}.
With streamers present, the voltage peak value was significantly reduced, possibly due to the series resistor limiting the current to the live electrode.

The actual voltage shape experienced by the test object is an important limitation in the simulations.
Since the fluid simulations were not linked to a model of the external circuit, the test object voltage could not be consistently coupled to the simulations, and we therefore used the prospective voltage for the simulations.
Several measurements were also made without the probe present, so the probe influence on the voltage is not accounted for.
However, the probe input impedance is high (\SI{600}{\mega\ohm}, \SI{15}{\pico\farad}), so the effect is likely minor.

To register light from the streamers, a Philips 56AVP photo-multiplier tube (PMT) with a spectral range of \SI{380}{\nano\meter} to \SI{680}{\nano\meter} was used.
Both the voltage probe and the PMT were connected to a \SI{1}{\giga\hertz} Tektronix DPO 4104 oscilloscope.

\Fref{fig:test_object_voltage} shows the PMT signal and measured voltage (both prospective and with streamers) over the test object.
The inset figure shows the events on the rising voltage front.
Signal transit times in cables and the internal PMT delay were compensated for in the post-processing.
The inception voltage as registered by the PMT showed a variation of some tens of ns, but the streamer event always occurred on the first rising voltage flank when using a $\SI{35}{\kilo\volt}$ charging voltage.
Typical instantaneous voltage values at inception were between $\num{20}$ and $\SI{30}{\kilo\volt}$.
When computing the inception voltage using the Townsend-Meek criterion, the resulting value was $U_{\textrm{inception}} \sim \SI{11}{\kilo\volt}$.
This deviation between experimental and calculated inception voltage is probably a result of waiting times for a starting electron and possibly non-zero surface charge on the dielectric.

Between each experiment, a grounded metal rod was guided over the surface to neutralize residual surface charge.
It is unlikely that this method removed all charge residues.
However, the test object will likely itself neutralize some charge during an impulse event:
As the voltage pulse decays, reverse discharges between the HV electrode and charged surface occur as can be seen on the PMT signal in \fref{fig:test_object_voltage} from around $\SI{40}{\micro\second}$.
These pulses neutralize some of the charge as shown in \cite{meyer_streamer_2019,meyer_streamer_2022}.
Although we can not exclude the influence of remnant surface charge in our experiments, the observed streamer behavior was highly reproducible.

\subsection{High speed imaging}
Different high-speed cameras were used to image the discharges:

\begin{enumerate}
\item A single-frame dual image intensifier camera, Lambert HiCAM 500, with an S20 photo-cathode, with max response wavelength 270 to 450 nm, full spectral range 200 to 800 nm and a 100 mm f2.8-8 CERCO UV lens (optimization range 250 to 410 nm, full range 220-900 nm).
  This camera was also used in \cite{meyer_streamer_2022}, but without a UV-transparent lens.
  The image intensifier was controlled with a gate pulse over a fiber optic link, see \fref{fig:schematic}.
  The on-time of the intensifier was varied between 10, 50 and \SI{100}{\nano\second}, and was recorded on an oscilloscope via a fiber optic link to the camera.
  This oscilloscope was then connected to the oscilloscope with voltage probe and PMT.
  Delays between oscilloscopes and in the camera itself were corrected in the post-processing to estimate camera exposure time relative to the voltage pulse and PMT.
  The intensifier gain and aperture of the lens was also varied, depending on the discharge brightness and intensifier on-time.

\item A 16-frame image intensifier camera, Specialized Imaging SIMX16 with an S25 photo-cathode (max response wavelength 500-850 nm, full range 200-900 nm) with a Nikkor 300 mm f2.8-22 lens.
  The camera splits incoming light in a total of 16 channels, providing up 16 frames of the same discharge.
  The frames were occasionally overlapped to increase the effective framerate up to 1 billion frames per second.
  Example images using this camera and description of the setup can be found in \cite{meyer_streamer_2020}.
  Since this camera system splits the incoming light, and does not have UV sensitivity, the image quality is poorer than for the HiCAM camera.
\end{enumerate}

The main viewing angle was from the front as shown in \fref{fig:profiles}b, with some variations in tilt to also capture depth variations of the discharge.
The images were post-processed by enhancing brightness and contrast and by overlaying an illuminated background picture of the setup.
The camera exposure relative to the voltage pulse was controlled with a digital delay generator.

\begin{table*}[h!t!b!]
  \centering
  \caption{\label{tab:air_reactions}.
    Air reaction mechanism for the simulation model.
    Note that two-body reactions have units of $\si{m}^3\si{s}^{-1}$ and three-body reactions have units of $\si{m}^6\si{s}^{-1}$.
    Reactions where the rate coefficient is given as $k = k(E,N)$ are computed with BOLSIG+\cite{Hagelaar2005a} and the PHELPS database \cite{PhelpsDatabase}.
    For these reactions the rate is given as a function of the reduced electric field $E^\prime = E/N$ (in units of Townsend).
    The electron temperature is obtained as $T_{\e} = 2\overline{\epsilon}/\left(3k_{\textrm{B}}\right)$ where $\overline{\epsilon}$ is the mean electron energy (computed using BOLSIG+).
    The gas temperature is $\Tg = 300\,\si{K}$.
  }
  \begin{indented}
  \item[]
    \centering
    \begin{tabular}{@{}lcll}
      \br
      Reaction \# & Reaction & Rate & Ref. \\
      \mr
      $1$ & $\e      + \Ntwo         \xrightarrow{k_1}   \NtwoP   + \e + \e       $ & $k_1 = k_1\left(E, N\right)$ & \cite{Hagelaar2005a, PhelpsDatabase} \\
      $2$ & $\e      + \Otwo         \xrightarrow{k_2}   \OtwoP   + \e + \e       $ & $k_2 = k_2\left(E, N\right)$ & \cite{Hagelaar2005a, PhelpsDatabase} \\
      $3$ & $\e      + \Otwo + \Otwo \xrightarrow{k_3}   \OtwoM   + \Otwo         $ & $k_3 = k_3\left(E, N\right)$ & \cite{Hagelaar2005a, PhelpsDatabase} \\
      $4$ & $\e      + \Otwo         \xrightarrow{k_4}   \OM      + \Oone         $ & $k_4 = k_4\left(E, N\right)$ & \cite{Hagelaar2005a, PhelpsDatabase} \\
      $5$ & $\OtwoM  + \Ntwo         \xrightarrow{k_5}   \Otwo    + \Ntwo + \e    $ & $1.13\times 10^{-25}$ & \cite{Kossyi1992} \\
      $6$ & $\OtwoM  + \Otwo         \xrightarrow{k_6}   \Otwo    + \Otwo + \e    $ & $2.2\times 10^{-24}$ & \cite{Kossyi1992} \\
      $7$ & $\OM     + \Ntwo         \xrightarrow{k_7}   \e       + \NtwoO        $ & $1.16\times 10^{-18}\exp\left[-\left(\frac{48.9}{11 + 2\Etd}\right)^2\right]$ & \cite{Pancheshnyi2013}\\
      $8$ & $\OM     + \Otwo         \xrightarrow{k_8}   \OtwoM   + \Oone         $ & $6.96\times 10^{-17}\exp\left[-\left(\frac{198}{5.6 + \Etd}\right)^2\right]$ & \cite{Pancheshnyi2013}\\
      $9$ & $\OM     + \Otwo + \M    \xrightarrow{k_9}   \OthreeM + \M            $ & $1.1\times 10^{-42}\exp\left[-\left(\frac{\Etd}{65}\right)^2\right]$ & \cite{Pancheshnyi2013}\\
      $10$ & $\NtwoP + \Ntwo + \M    \xrightarrow{k_{10}} \NfourP  + \M            $ & $5\times 10^{-41}$ & \cite{Aleksandrov1999}\\
      $11$ & $\OtwoP + \Otwo + \M    \xrightarrow{k_{11}} \OfourP  + \M            $ & $2.4\times 10^{-42}$ & \cite{Aleksandrov1999}\\
      $12$ & $\e     + \OfourP       \xrightarrow{k_{12}} \Otwo    + \Otwo         $ & $1.4\times 10^{-12}\sqrt{\frac{T_{\e}}{\Tg}}$ & \cite{Kossyi1992}\\
      $13$ & $\e     + \NfourP       \xrightarrow{k_{13}} \Ntwo    + \Ntwo         $ & $2\times 10^{-12}\sqrt{\frac{T_{\e}}{\Tg}}$ & \cite{Kossyi1992}\\
      $14$ & $\e     + \Ntwo         \xrightarrow{k_{14}} \e       + \Ntwo + \gamma_j$ & See text. & \cite{Hagelaar2005a, PhelpsDatabase} \\
      $15$ & $\gamma_j + \Otwo         \rightarrow         \e       + \OtwoP        $ & See text  & \cite{1982TepVT..20..423Z}\\
      \br
    \end{tabular}
  \end{indented}
\end{table*}

\section{Numerical methods}
\label{sec:computer_simulations}

\subsection{Streamer dynamics model equations}
The numerical simulations were performed using a Cartesian 2D fluid model.
Fluid models are often used for simulation of streamer discharges in air, see e.g. \cite{viegas_investigation_2018, Soloviev2009, Morrow1997, bagheri_comparison_2018, Babaeva2018, soloviev_mechanism_2014}.
The equations of motion are:

\begin{eqnarray}
  \label{eq:species}
  &\frac{\partial n}{\partial t} = -\nabla\cdot\left(\bm{v}n - D\nabla n\right) + S, \\
  \label{eq:sigma}
  &\frac{\partial \sigma}{\partial t} = J_\sigma, \\
  \label{eq:poisson}
  &\nabla\cdot\left(\epsilon_r\bme\right) = \frac{\rho}{\epsilon_0}, \\
  \label{eq:rte}
  &\left[\nabla^2 - \left(p_{\Otwo}\lambda_j\right)^2\right]\Psi_j = -\left(A_j p_{\Otwo}^2 \frac{p_q}{p_q + p}\xi\nu\right)k_1 n_{\e} n_{\Ntwo}.
\end{eqnarray}
Here, $n$ is the particle number density, $\bm{v} = \pm \mu(E) \bm{E}$ is the drift velocity for positively ($+$) and negatively ($-$) charged species, where $\mu$ is the species mobility and $\bm{E} = -\nabla\Phi$ is the electric field.
Furthermore, $D$ is the diffusion coefficient, and $S$ indicates source terms for the various species.
These are discussed further below.
The surface charge density is given by $\sigma$ where $J_\sigma$ is the charge flux onto the dielectric surface.
The relative permittivity of the dielectric is $\epsilon_r = 3$ and $\rho$ is the space charge.
Finally, \eref{eq:rte} describes photoionization in air; we elaborate on this equation below.

\subsection{Transport data}
For the plasma chemistry we solve for the following 8 species: $\textrm{e}$, $\textrm{N}_2^+$, $\textrm{O}_2^+$, $\textrm{O}_2^-$, $\textrm{O}^-$, $\textrm{O}_3^-$, $\text{N}_4^+$, $\text{O}_4^+$.
The electron mobility and diffusion coefficients are computed using BOLSIG+ \cite{Hagelaar2005a} with the PHELPS database \cite{PhelpsDatabase}, while for the ions we take $\mu = 2\times 10^{-4}\,\si{m}^2/(\si{Vs})$ and $D = 0$.
For the source terms we use the reactions given in \tref{tab:air_reactions}.

For the photoionization reactions (reactions \#14 and \#15), we consider the \textcite{Bourdon2007} Helmholtz reconstruction where the photoionization reaction is the sum of three contributions:

\begin{equation}
  S_{\textrm{14}} = \sum_{j=1}^3 \Psi_j.
\end{equation}
Here, $\Psi_j$ are the solutions to \eref{eq:rte} where $p_q = 40\,\si{mbar}$ is a quenching pressure (and $p = 1\,\si{bar}$ is the gas pressure), $\xi$ is the photoionization efficient and $\nu$ is the relative excitation probability.
That is, $\nu k_1n_{\e} n_{\Ntwo}$ is the electron impact excitation rate of molecular nitrogen.
Spontaneous emission from these can lead to photoionization of molecular oxygen.
In this paper we take $\xi\nu = 0.06$.
The $\lambda$ and $A$ coefficients are given in \tref{tab:photoionization}.

\begin{table}[h!t!b!]
  \caption{
    \label{tab:photoionization}
    Model parameters for the Helmholtz reconstruction.
  }
  \begin{indented}
  \item[]
    \centering
    \begin{tabular}{@{}lll}
      \br
      $j$ & $\lambda\,\left(\si{m}^{-2}\si{Pa}^{-2}\right)$ & $A\,\left(\si{m}^{-1}\si{Pa}^{-1}\right)$ \\
      \mr
      $1$ & $0.415$  & $1.12\times 10^{-4}$ \\
      $2$ & $0.1095$ & $2.87\times 10^{-3}$ \\
      $3$ & $0.6675$ & $0.275$ \\
      \br
    \end{tabular}
  \end{indented}
\end{table}

\subsection{Discretization}
The equations are discretized in space using finite volumes on hierarchical Cartesian grids.
We use an embedded boundary (EB) formalism for representing the electrode and dielectrics.
EBs add substantial discretization complexity which is not discussed in detail here, see e.g. \textcite{ebchombo} or \textcite{Marskar2019a}.
The simulations are performed in planar 2D, with a finest grid resolution of  $2.44\,\mu\si{m}$.
For discretization in time we use a Godunov splitting method as in \textcite{marskar_3d_2020}.
The only difference between the current discretization and \textcite{marskar_3d_2020} is that we here use a semi-implicit coupling to the electric field.

For the rectangular corrugated surface we found that the simulations required extremely fine resolutions when the streamer temporarily halts in the corrugation spaces.
In order to stabilize the simulations we used the upwinding suggestions from \textcite{Villa2014} where the source term is computed using the upwinded value of the electron density.
In practice, we enforce this by using the face-centered states that are available from the hyperbolic discretization, which thus also includes the effects of slope limiters.
For consistency, this modification is done for all the simulations and profiles.

For boundary conditions, the left, right, and bottom edges in the simulation domain are grounded.
On the top edge we impose a homogeneous Neumann boundary condition $\partial_y\Phi = 0$, and on the electrode we impose the time-varying prospective voltage shown in \fref{fig:test_object_voltage}.
On the dielectric surface we incorporate the following boundary conditions into the discretization of the Poisson equation:

\begin{equation}
  \label{eq:jump_bc}
  \epsilon_1\partial_{n_1}\Phi + \epsilon_2\partial_{n_2}\Phi = \frac{\sigma}{\epsilon_0},
\end{equation}
where $\sigma$ is the surface charge density and $\mathbf{n}_1 = -\mathbf{n}_2$ are unit normals that points away from the surface and into the plasma.
The permittivies of the gas and dielectric phases are here represented by $\epsilon_1 = 1$ and $\epsilon_2 = 3$.

\subsection{Initial conditions}
\begin{figure}[h!t!b!]
  \centering
  \includegraphics{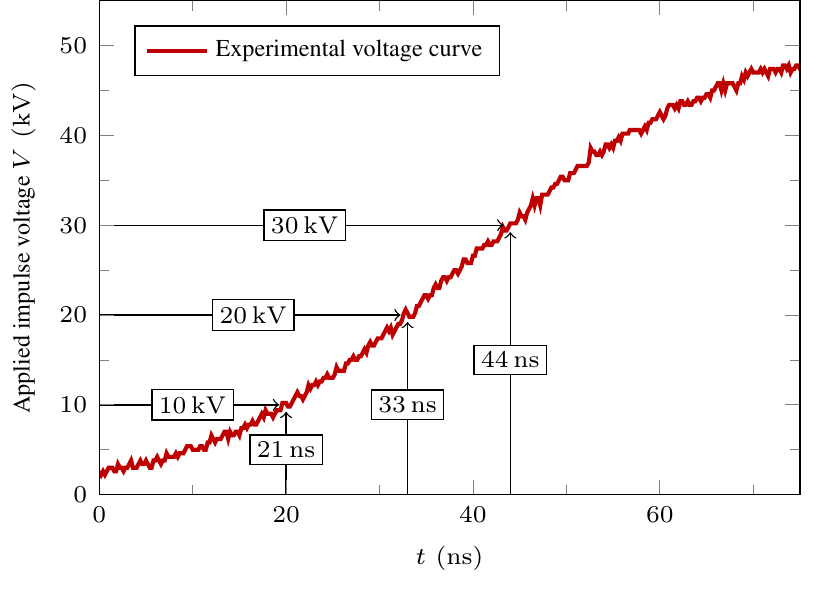}
  \caption{Experimental voltage curve (rising flank only) and time lags used in the streamer simulations.}
  \label{fig:streamer_inception_time_lag}
\end{figure}

In the experiments we observed that the streamers started on the rising voltage flank, typically between $20\,\si{kV}$ and $30\,\si{kV}$ instantaneous voltage values.
To account for experimental uncertainties as well as natural variations in the statistical time lag in the simulations, we run simulations that start at voltages of $10\,\si{kV}, 20\,\si{kV}$, and $30\,\si{kV}$ on the voltage curves.
These values represent the lowest possible inception voltage ($10\,\si{kV}$) and the typical experimentally observed inception voltages ($20\,\si{kV}$ and $30\,\si{kV}$).
The corresponding time lags are shown in \fref{fig:streamer_inception_time_lag}.
We do not expect the simulations at $10\,\si{kV}$ to hold much relevance since, at this voltage, the critical volume around the electrode was extremely small and the probability for inception was therefore also very low.

In all cases we initialize the simulation using a small Gaussian seed of electron-ion pairs near the electrode tip, given by

\begin{equation}
  n_\e\left(\mathbf{x}, t\right) = n_0\exp\left[-\frac{\left(\bm{x}-\bm{x}_0\right)^2}{2R^2}\right],
\end{equation}
where $n_0 = 10^{16}\,\si{m}^{-3}$, $\bm{x}_0$ is the electrode tip and $R=100\,\mu\si{m}$.
For the ions we used $n_{\OtwoP} = 0.2n_\e$ and $n_{\NtwoP} = 0.8n_\e$.

\begin{figure*}[h!t!b!]
  \centering
  \includegraphics{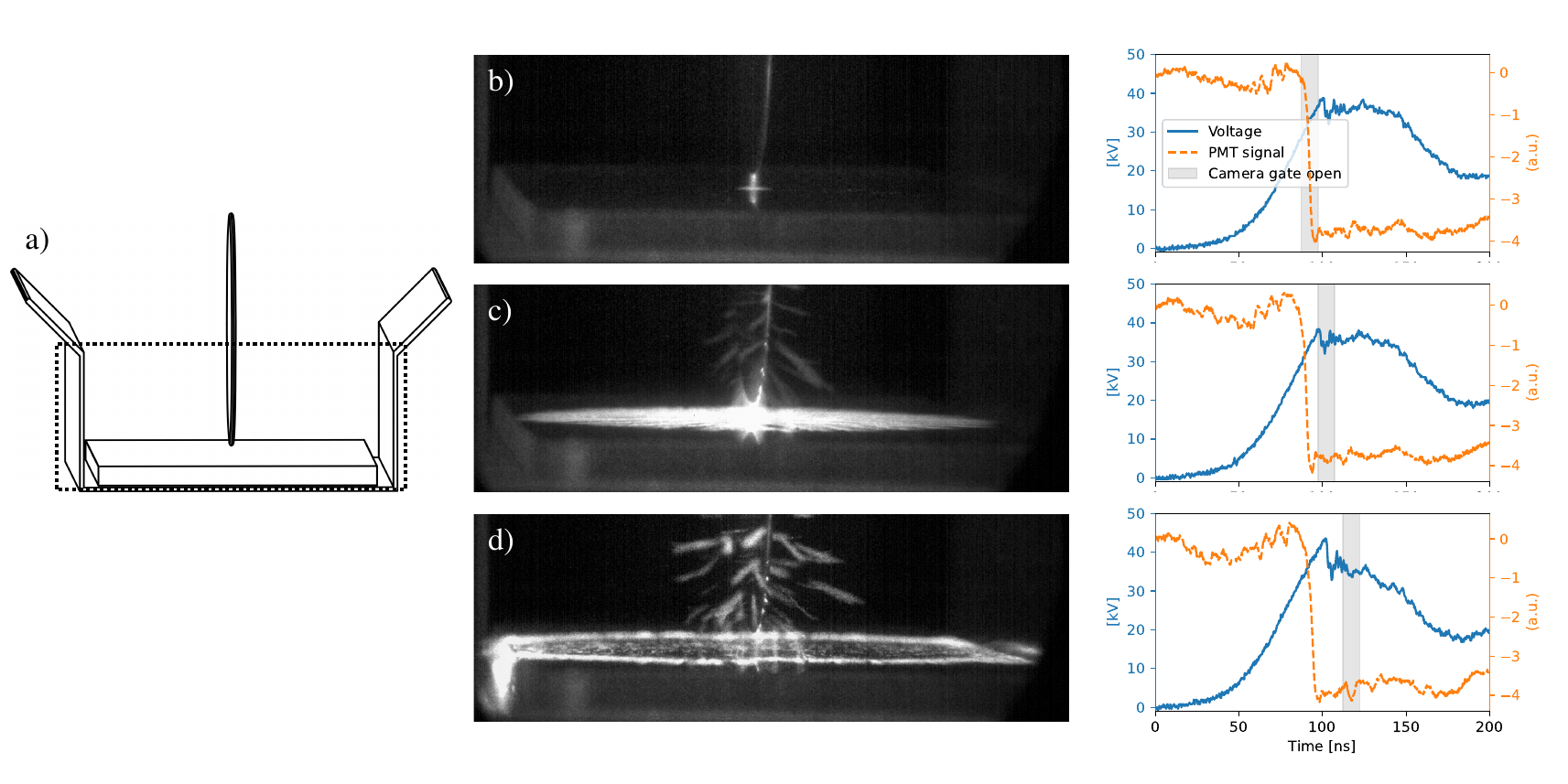}
  \caption{Example images of streamers on a smooth dielectric surface with corresponding plots of voltage, PMT signal and camera exposure time.
    Each image is a taken during a different experiment as the HiCAM is a single-frame camera.
    a) Sketch of image view.
    b) \SI{10}{\nano\second} camera exposure during streamer inception.
    c) \SI{10}{\nano\second} camera exposure after inception.
    d) \SI{10}{\nano\second} camera exposure towards the last stage of surface streamer propagation.
  }
  \label{fig:hicam_flat}
\end{figure*}

\begin{figure*}[h!t!b!]
  \centering
  \includegraphics{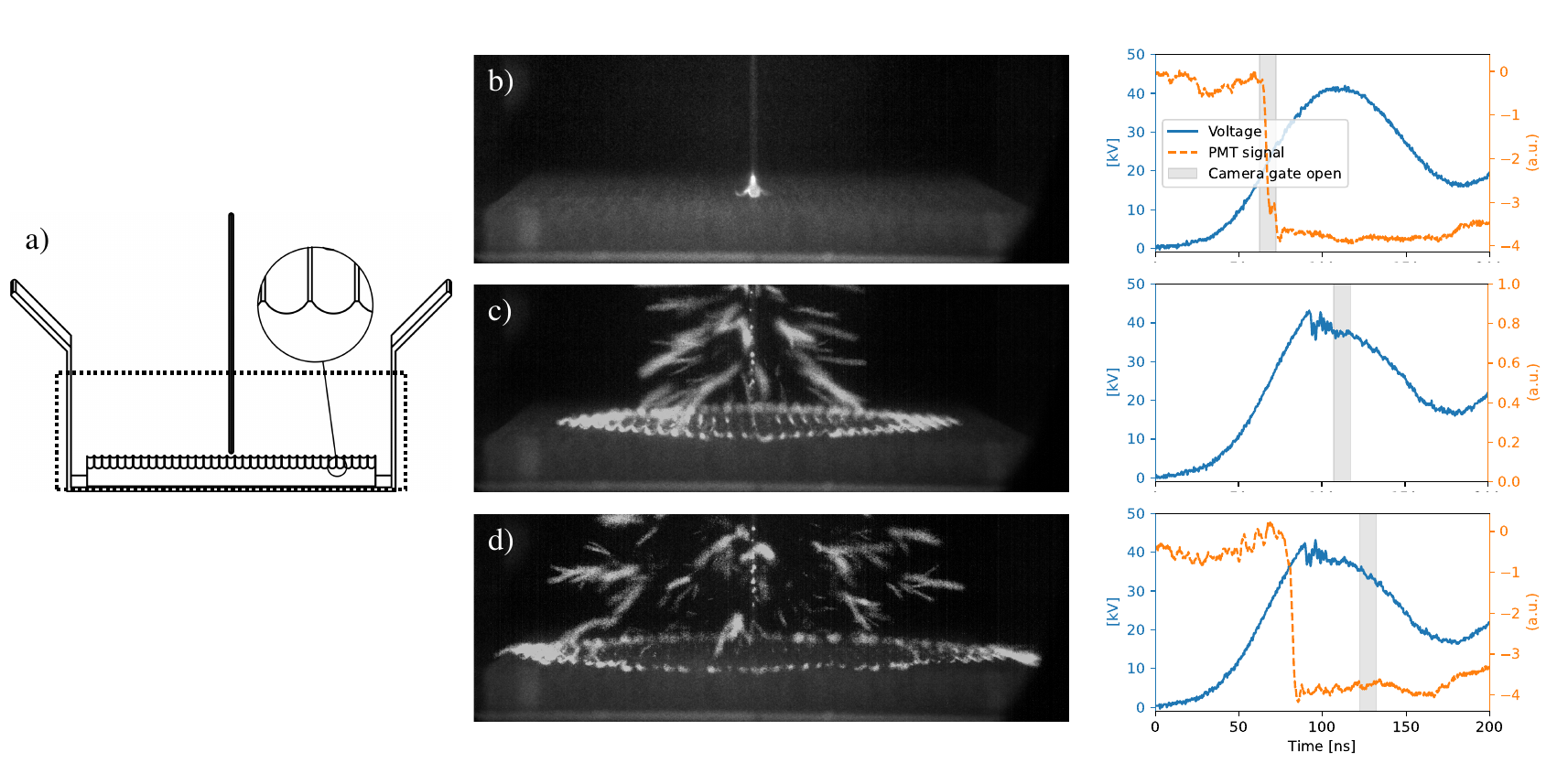}
  \caption{Streamers on a semi-circular profiled dielectric surface with corresponding plots of voltage, PMT signal and camera exposure time.
    Each image is a taken during a different experiment as the HiCAM is a single-frame camera.    
    a) Sketch of image view.
    b) \SI{10}{\nano\second} camera exposure image at inception.
    c) \SI{10}{\nano\second} exposure image during propagation. The PMT was accidentally switched off during this experiment, so no PMT signal is shown.
    d) \SI{10}{\nano\second} exposure when the streamer reaches the end of the plate.
  }
  \label{fig:hicam_circular}
\end{figure*}

\begin{figure*}[h!t!b!]
  \centering
  \includegraphics{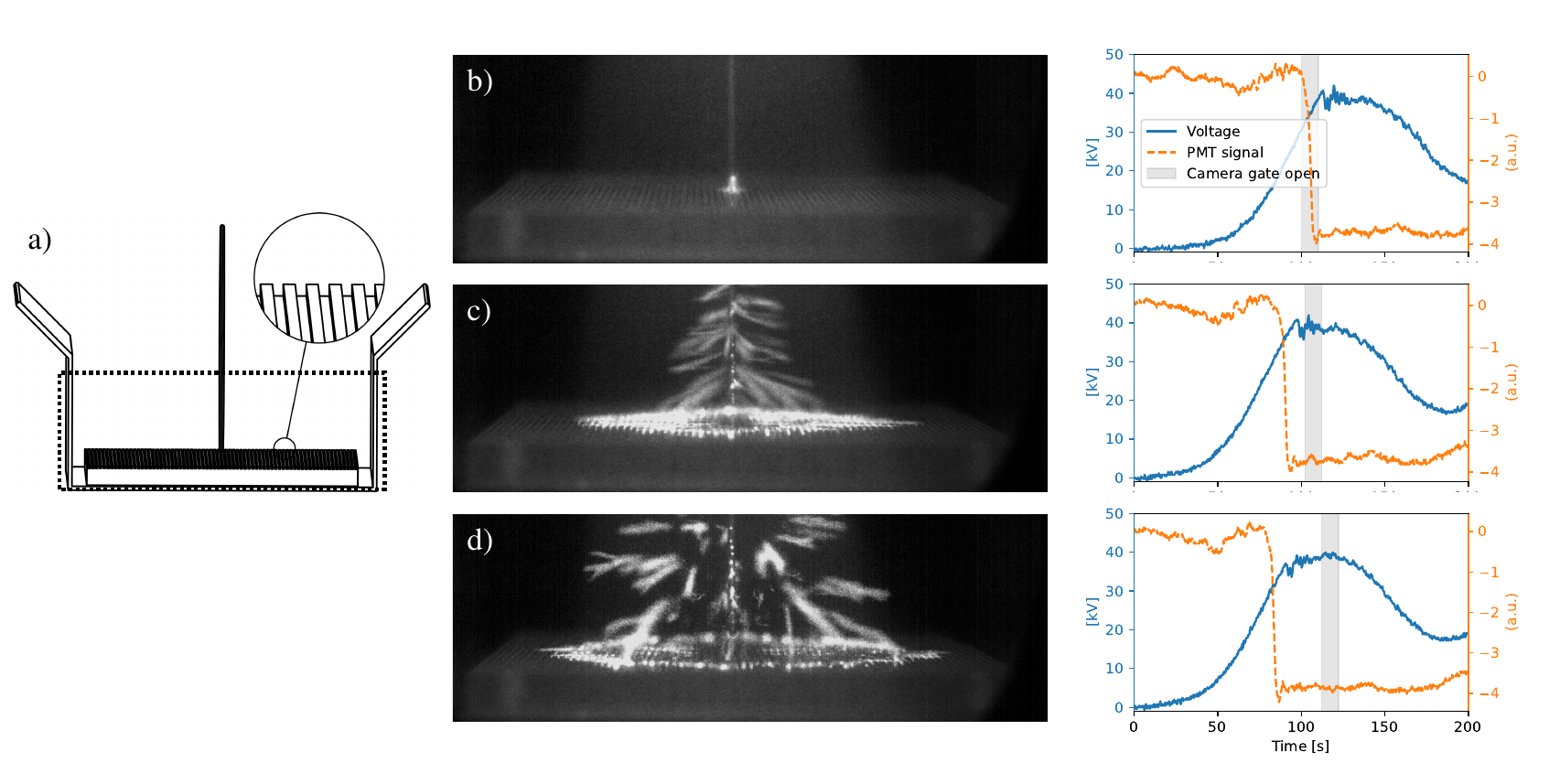}
  \caption{Example images of streamers on a rectangular profiled dielectric surface with corresponding plots of voltage, PMT signal and camera exposure time.
    Each image is a taken during a different experiment as the HiCAM is a single-frame camera.        
    a) Sketch of image view.
    b) \SI{10}{\nano\second} camera exposure image at inception.
    c) \SI{10}{\nano\second} exposure image during propagation.
    d) \SI{10}{\nano\second} exposure during later-stage streamer propagation.
  }
  \label{fig:hicam_square}
\end{figure*}

\begin{figure*}[h!t!b!]
  \centering
  \includegraphics{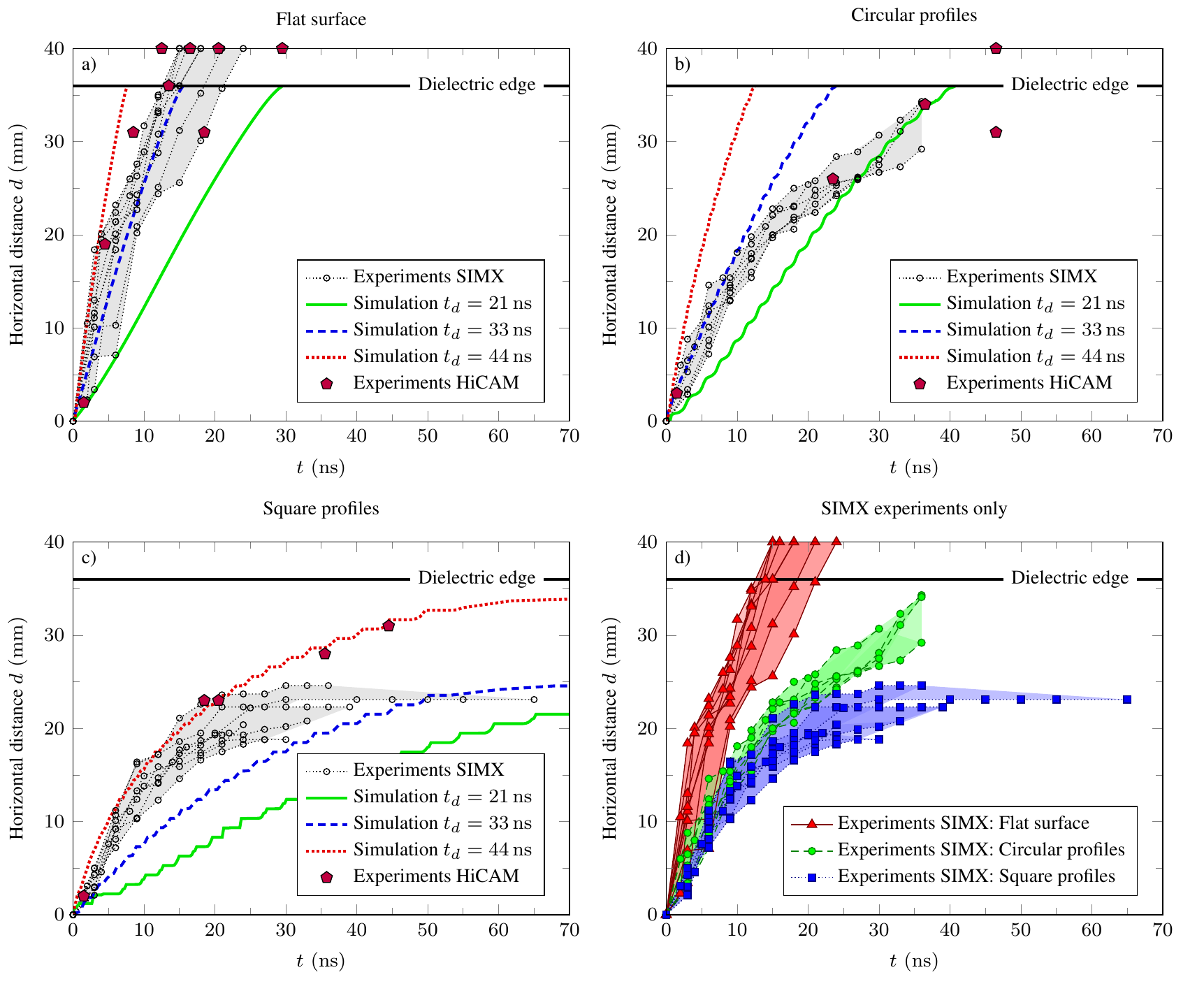}
  \caption{
    Travel curves for experiments and simulations. Results from both the multi-frame SIMX camera and the single-frame, but UV-sensitive HiCAM camera are shown.
    a) Flat surface, experiments and simulations.
    b) Semi-circular surface corrugations, experiments and simulations.
    c) Rectangular surface corrugation, experiments and simulations.
    d) Comparison of experimental results for the three surfaces.
    For the SIMX camera experiments, $t=0$ indicates is the first non-empty image, while for the simulations $t=0$ corresponds to when the streamer started propagating along the dielectric surface.
    For the HiCAM experiments, $t=0$ was estimated from the PMT signal, and then adjusted such that $t=0$ was consistent with the SIMX experiments for distances $d\lesssim\SI{5}{\milli\meter}$.
    The color shading highlights the extent of the SIMX experimental results.
  }
  \label{fig:velocities}
\end{figure*}

\section{Results and discussion}
\label{sec:results}
\subsection{Streamer morphology}
Example images of streamer propagation with corresponding plots of voltages, PMT signals, and camera exposure time windows are shown in \fref{fig:hicam_flat}, \fref{fig:hicam_circular} and \fref{fig:hicam_square}.
The images are 10 ns single-shot frames taken with the HiCAM camera, and each image comes from a different experiment.

\Fref{fig:hicam_flat}b)-d) show \SI{10}{\nano\second} snapshots of streamers propagating over a flat surface.
\Fref{fig:hicam_flat}b) is taken directly after inception, where the streamer has just crossed the air gap, hit the dielectric surface and started propagating along it.
The lower part of the streamer light seen in this image is likely a reflection from the transparent polycarbonate surface of the air gap streamer.
The surface streamer is comparatively planar, behaving almost like an ionization wave.
However, some branching can be seen in \fref{fig:hicam_flat}c) as the streamer propagates along the surface.
In \fref{fig:hicam_flat}c) and d), air-borne streamers from other parts of the HV electrode start propagating.
These do not catch up with the surface streamer, which reach the ground plane in \fref{fig:hicam_flat}d).

\Fref{fig:hicam_circular}b)-d) show \SI{10}{\nano\second} snapshots of streamers propagating over a semi-circular profiled surface, with corresponding plots of voltage, PMT signal and camera exposure window, analogous to \fref{fig:hicam_flat}b)-d).
For this surface we found that the streamer descends into the corrugations as predicted by simulations in \cite{meyer_streamer_2020}, see \fref{fig:hicam_circular}b), or \cite{meyer_streamer_2022} for a different viewing angle.
Another observed effect is the illumination of peaks on the surface profile in the streamer wake, see \fref{fig:hicam_circular}c).
Unlike the flat surface in \fref{fig:hicam_flat}, the airborne streamers in \fref{fig:hicam_circular} almost catch up with the surface streamer.

\Fref{fig:hicam_square}b)-d) show \SI{10}{\nano\second} snapshots of streamers propagating over a rectangular profiled surface, with corresponding plots of voltage, PMT signal and camera exposure window.
Although not as clear as on the semi-circular surface, the surface streamer does appear to propagate down into the corrugations.
Additionally, illumination of surface profile peaks are again observed in the streamer wake (see \fref{fig:hicam_square}d).
From long-exposure images shown in \cite{meyer_streamer_2022} we could see that the surface streamer stagnates on rectangular corrugations, and does not reach the ground plane. 

\subsection{Travel curves}
To estimate the streamer velocity we used the SIMX-camera to estimate the location of the streamer head.
Since the SIMX camera has 16 channels for single-exposure images, we had up to 16 frames per discharge that we could use for estimating the instantaneous streamer head positions.
The streamer horizontal propagation distance in each individual frame was estimated visually from the images with the method described in \cite{meyer_streamer_2020}.
Only image series that began with an empty frame were used in the analysis. The temporal uncertainty of the discharge inception is therefore equal to the frame duration, which was varied between $2\,\textrm{ns}$ and $10\,\textrm{ns}$.
We extracted five travel curves for each surface, which are plotted together with the simulation results in \fref{fig:velocities}.

Although the SIMX camera provided travel curves with satisfactory temporal resolution, the level of detail in the SIMX images was much lower than for the HiCAM.
When analyzing the frames we found that light emission from the surface streamer is strongest close to the electrode, and that it weakens as the streamer propagates along the surface.
Comparing SIMX images with the HiCAM images showed that, at least for the rectangular surface profile, some of the empty frames from the SIMX camera contained surface streamers that were too faint to be visible on the SIMX camera.
To provide extra data points outside the available range of the SIMX camera, the mean streamer velocity was also estimated from HiCAM images.
We measured the surface streamer front position in each image, and estimated the propagation time as the time between the activation of the PMT (which indicates inception) and the closing of the camera gate.
These data points are plotted as filled pentagons in \fref{fig:velocities}.
We point out that the PMT signal has a finite rise time and evaluation of the inception time from this signal has an uncertainty of at least \SI{5}{\nano\second}.
To adjust the accuracy of these measurements we chose the inception time such that the streamer positions for the HiCAM and SIMX cameras were consistent for propagation distances $d\lesssim\SI{5}{\milli\meter}$.
Correspondingly, all the simulation data in \fref{fig:velocities} have a common reference time ($t=0$) where the streamers start propagating along the surface.

In general, we found that the streamers propagated faster over the flat dielectric than over the profiled surfaces.
In the experiments with a flat surface the surface streamers propagated with a relatively constant velocity of about $2\,\si{mm}/\si{ns}$, while for the semi-circular profile the mean horizontal velocity was around $1.2\,\si{mm}/\si{ns}$.
For the rectangular-cut surface, the surface streamers did not cross the dielectric and were substantially slower, which is in line with experimental observations reported in \cite{meyer_streamer_2022}.

\Fref{fig:velocities} also shows the corresponding velocities obtained from computer simulations.
The simulations were performed for streamers starting after time delays $t_d = \SI{21}{\nano\second}$, $t_d=\SI{33}{\nano\second}$, and $t_d=\SI{44}{\nano\second}$ to simulate the effect of a first electron time-lag.
The curves in \fref{fig:velocities} show the position of the maximum value of the photoionization source term ($\propto k_1n_{\e}$) for these three simulations.
The fluctuation on the simulated travel curves for the profiled surfaces correspond to changes in horizontal streamer velocity as the streamers travel into and out of the surface corrugations.

\begin{figure*}[h!t!b!]
  \centering
  \includegraphics{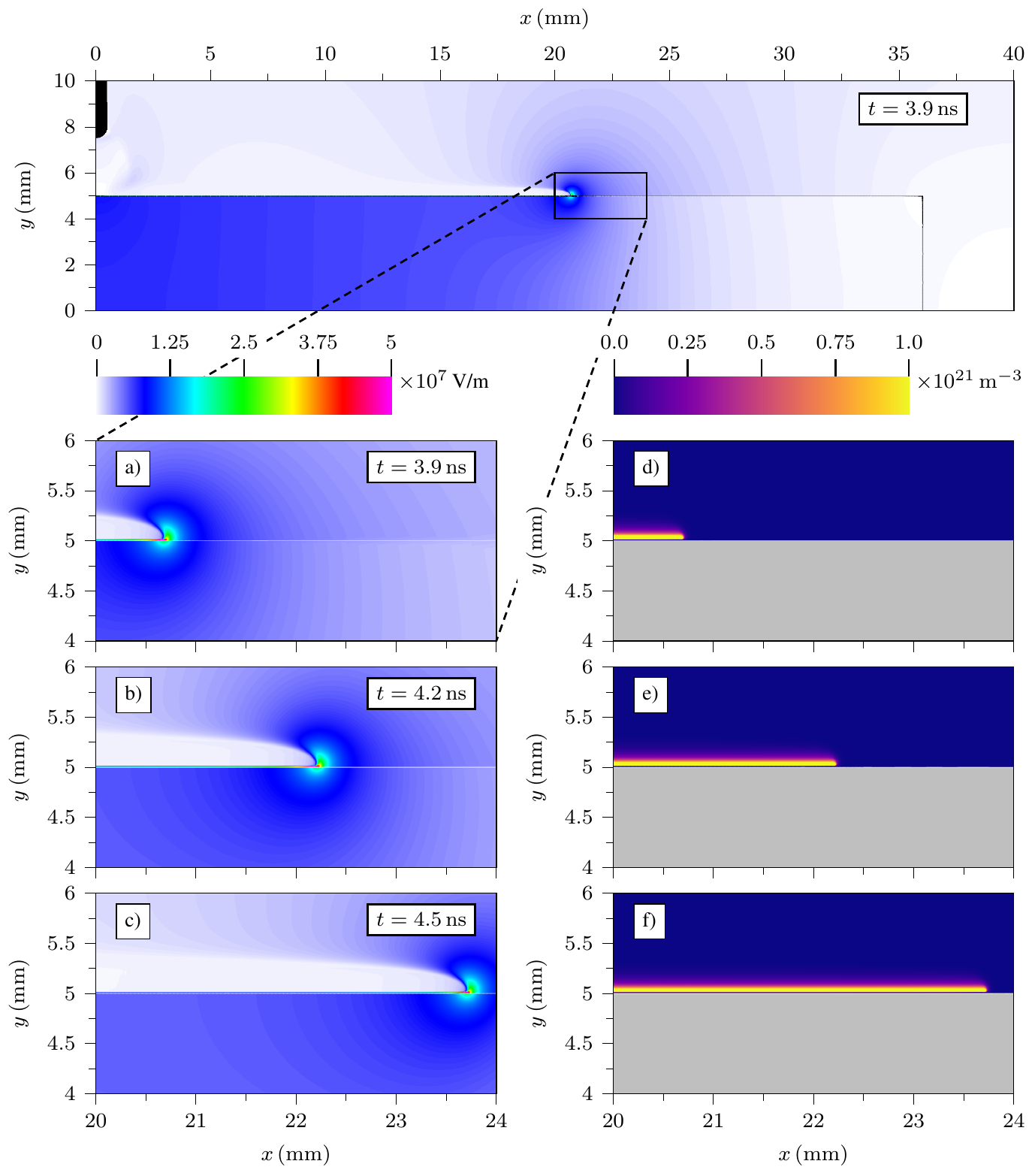}
  \caption{Evolution snapshots for propagation over the flat surface for a time delay $t_d=44\,\si{ns}$.
    Top panel: Field distribution around the dielectric surface. 
    a), b), and c): Snapshots of the electric field magnitude $\left|\bm{E}\right|$.
    d), e), and f): Electron density, clamped to a color map $n_{\e} \in \left[0,10^{21}\right]\,\si{m}^{-3}$.
    }
  \label{fig:flat_evolution_30kv}
\end{figure*}

\begin{figure*}[h!t!b!]
  \centering
  \includegraphics{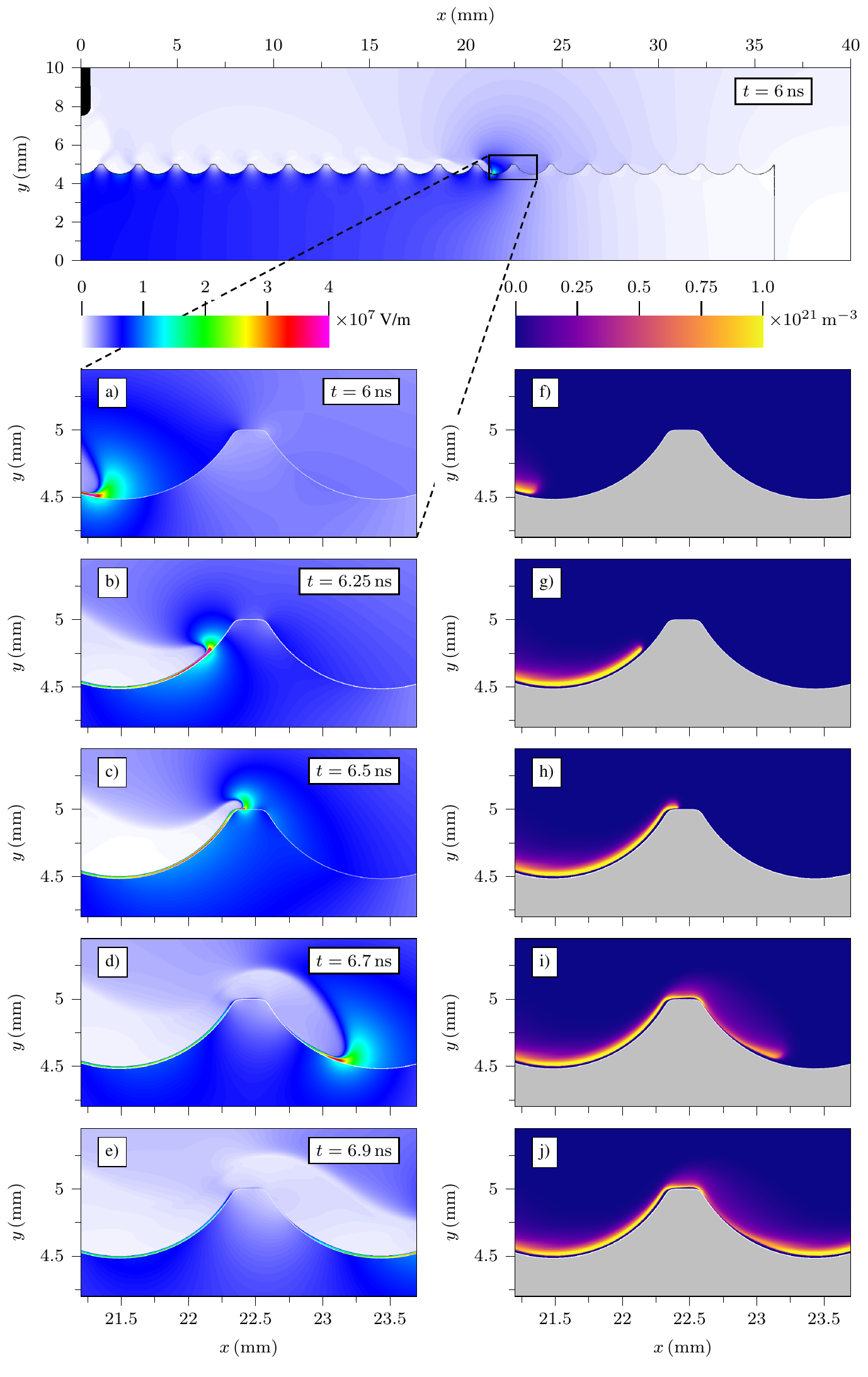}
  \caption{Evolution snapshots for the semi-circular surface profile for a time delay $t_d=44\,\si{ns}$.
    Top panel: Field distribution around the dielectric surface.     
    a) through e): Snapshots of the electric field magnitude $\left|\bm{E}\right|$.
    f) through j): Electron density, clamped to a color map $n_{\e} \in \left[0,10^{21}\right]\,\si{m}^{-3}$.
    }
  \label{fig:circle_evolution_30kv}
\end{figure*}

\begin{figure*}[h!t!b!]
  \centering
  \includegraphics{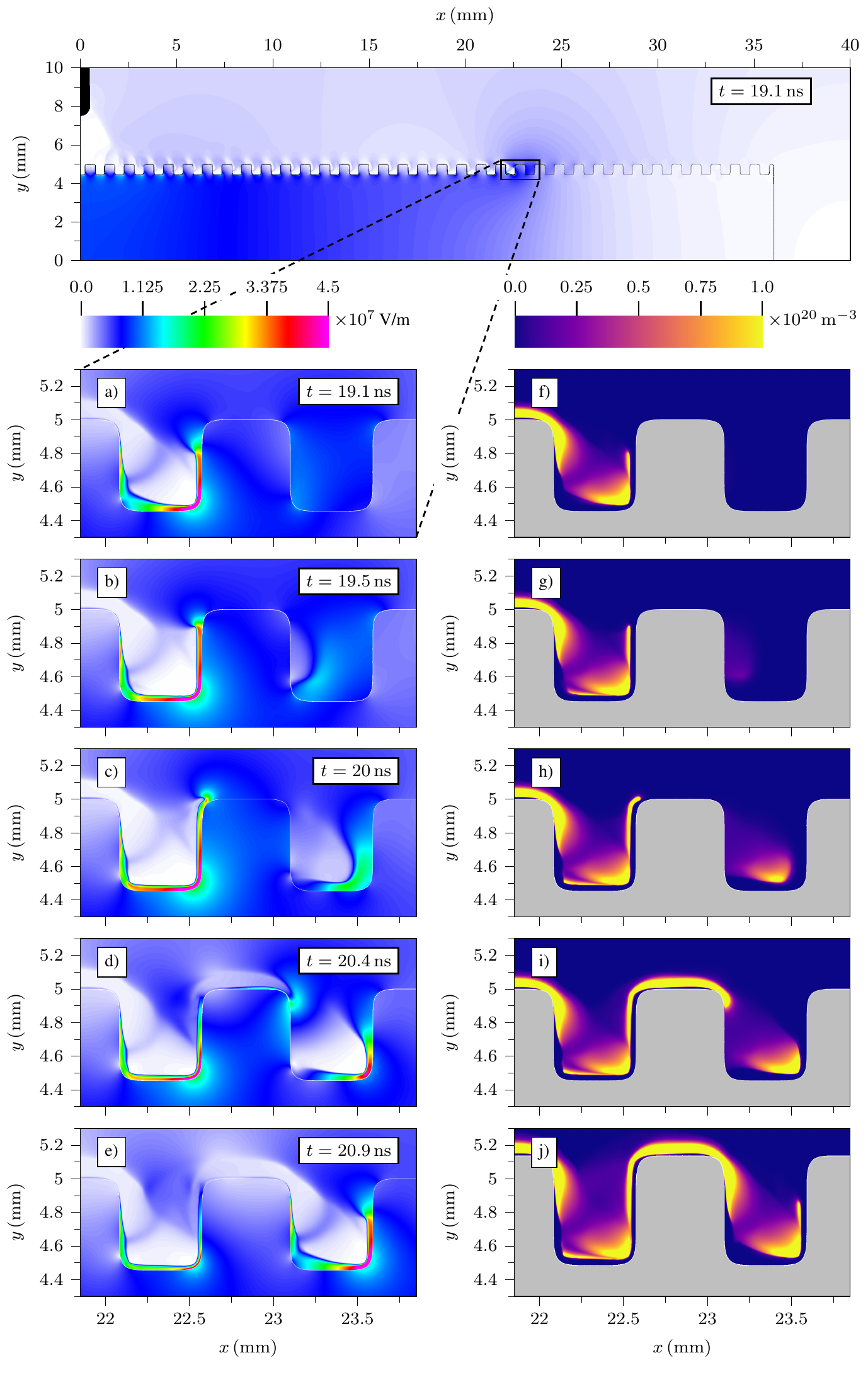}
  \caption{Evolution snapshots for the rectangular surface profile for a time delay $t_d=44\,\si{ns}$.
    Top panel: Field distribution around the dielectric surface.         
    a) through e): Snapshots of the electric field magnitude $\left|\bm{E}\right|$.
    f) through j): Electron density, clamped to a color map $n_{\e} \in \left[0,10^{20}\right]\,\si{m}^{-3}$.
    }
  \label{fig:square_evolution_30kv}
\end{figure*}

\subsection{Propagation mechanism}
Next, we discuss the propagation mechanism of the positive streamers over the three surfaces.
\Fref{fig:flat_evolution_30kv}, \fref{fig:circle_evolution_30kv}, and \fref{fig:square_evolution_30kv} show snapshots of the electric field and the plasma density $n_\e$ for all three surfaces.
The data in the figures are taken from the simulation runs that started on $30\,\si{kV}$, i.e. a statistical time lag of $44\,\si{ns}$ relative to the voltage curve. 
However, we found that the propagation mechanisms were the same for all three time lags that were investigated for each surface.
The largest difference between the different time-lags were the streamer velocities and range, and not overall morphologies.
For this reason we do not show corresponding plots for the simulations starting at $t_d=21\,\si{ns}$ and $ t_d=33\,\si{ns}$.
We point out that the time labels in the figures indicate the time from the appearance of the starting electron.
The temporal location relative to the voltage curve can be inferred from the provided time lags in \fref{fig:streamer_inception_time_lag}.

For the flat surface we find that the streamer propagates in a uniform manner over the surface.
As observed in many other computer simululations \cite{Soloviev2014, Marskar2019a, meyer_streamer_2020, li_computational_2020, meyer_streamer_2022}, the streamer is accompanied by a cathode sheath near the dielectric surface.
In general, the positive streamer is still driven by ionization processes in the gas, where photoionization provides the majority of seed electrons in front of it.
For further details regarding the dynamics of positive streamers over dielectric surfaces, see \textcite{li_computational_2020}. 

For the semi-circular profile surface we identify the same propagation mechanism as in \cite{meyer_streamer_2020}.
The streamer hugs to the surface for the entire propagation, and does not detach from the surface.
In addition, propagation from the profile peak and downwards into the pore is generally faster than from the bottom of the pore and towards the top of the profile.
For further details on the propagation, see \cite{meyer_streamer_2020} and \textcite{wang_effect_2021}.

\begin{figure}[h!t!b!]
  \centering
  \includegraphics{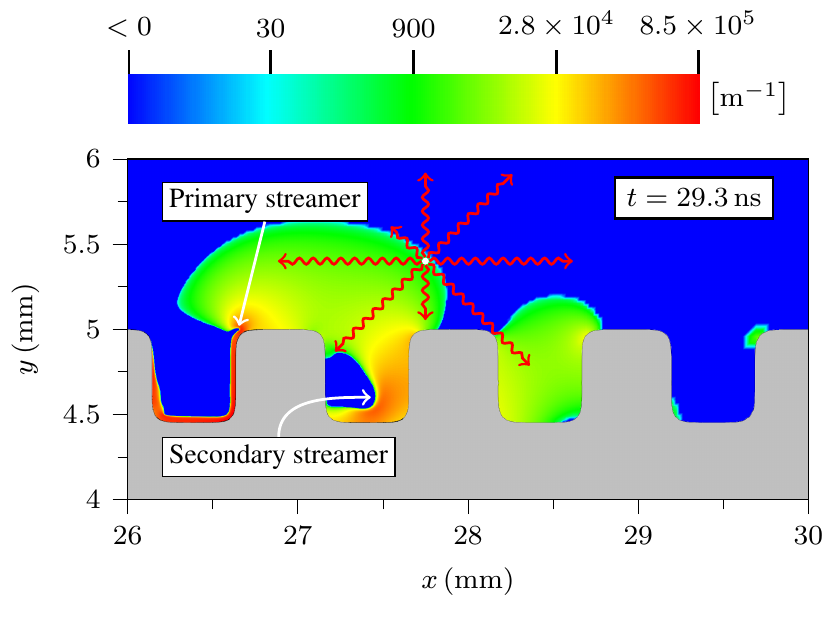}
  \caption{Feasibility sketch of generating starting electrons in pore spaces.
    The color-coded data shows the ionization zones ahead of the streamer, i.e. $\log \left[\max\left(\alpha-\eta, 0\right)\right]$.
    Squiggly lines indicate potential photon paths from de-excitation of $N_2$ triplet states \cite{Stephens2018} at the edge of the ionization zone.
    The data is taken from the simulation with $t_d=44\,\si{ns}$. 
  }
  \label{fig:square_ionization_zones}
\end{figure}

The propagation mechanism for the rectangular-cut surface differs substanstially from the other two surfaces we investigated.
\Fref{fig:square_evolution_30kv}a)-e) show the evolution of the electric field as the streamer propagates from one pore to the next, and \fref{fig:square_evolution_30kv}f)-j) show the corresponding plasma density.
From these frames we discern two main propagation mechanisms that operate in parallel:
\begin{itemize}
\item A primary streamer that propagates upwards out of the initial pore.
\item Inception of a secondary streamer, or partial discharge, in the neighboring pore. 
\end{itemize}
In these simulations we find that the secondary streamer fills the pore \emph{before} the main streamer crosses the profile peak and enters the next pore.
This is particularly visible in \fref{fig:square_evolution_30kv}c) and h).
The feasibility of this mechanism depends on the existence of a free electron in the neighboring pore.
Unfortunately, our numerical methodology does not allow us to test this in a conclusive way.
Our radiative transfer approach is essentially based on a diffusion equation (strictly speaking, an Eddington approximation) \cite{Marskar2019a}.
Even when the appropriate absorbing boundary conditions are used \cite{Larsen2002}, the Eddington approximation does not accurately capture shadows.
Ionizing photons can correspondingly ''leak'' around boundaries such that a starting electron, or at least a fraction of it, is always available. 
In our present model, however, we can still provide a few rough evaluations of the feasibility of this mechanism.
\Fref{fig:square_ionization_zones} shows the effective ionization coefficient $\alpha - \eta$ for the rectangular profiled surface, clamped to a positive range (i.e., showing the ionization zones).
As the primary streamer climbs out of the pore, and the secondary streamer fills it, photo-electrons will appear in the avalanche zone ahead of the primary streamer.
These electrons also generate new avalanches and ionizing photons.
The ionizing photons are emitted isotropically and their mean free path path length is on the order of \SI{500}{\micro\meter} \cite{Stephens2018}.
Rough geometric evaluations then show that the next pore space is neither completely shielded or too far away to be reached by ionizing photons.
However, we can only conclusively answer this by using discrete photons as in \cite{marskar_3d_2020}, or with a full particle method.

\subsection{Modeling uncertainties}
\Fref{fig:velocities} shows that velocity trends in both simulations and experiments are in relatively good quantitative agreement.
However, exact agreement is not expected since there are a number of uncertainties that can affect the computer simulations:

\begin{itemize}
\item The simulations are done in Cartesian 2D, but the experiments also exhibit a 3D morphology.
\item Streamer branches that start higher up on the disk electrode could affect the surface streamer, but this is not accounted for in the simulations.
\item The presence of starting electrons in the various pore spaces is not known.
\item In the experiments, the discharge affects the voltage over the test object.
\end{itemize}

Despite these uncertainties, the agreement between simulations and experiments is quite good.
Improvements to our modeling will most likely mandate a fully 3D approach.
This is particularly the case if one wants to include the electric circuit into the model, or the secondary streamer branches that initiate from the electrode.

\section{Conclusions}
\label{sec:conclusions}
In this work, positive streamer propagation over different surface profiles has been investigated.
We have performed an analysis consisting of

\begin{itemize}
\item High-speed imaging using two different camera systems.
\item Fluid simulations in 2D planar coordinates and direct comparison with experiments.
\end{itemize}

The high-speed imaging showed that we could, at least without introducing drastic modeling errors, approximate the discharges as Cartesian plane waves.
We used the imaging to estimate the streamer velocities and ranges for all surfaces, and we observed that the streamers retained their surface-hugging nature also when propagating down into surface corrugations.
The streamer discharges were impeded by the surface profiles, where the rectangular cut corrugations restricted the streamer range the most.
This surface also had the largest surface area, which could partially explain the result.

We also performed computer simulations which gave comparatively good agreement with experiments.
Presumably, part of this agreement arises from the fact that the surface streamers could be approximated as plane waves.
Surprisingly, for the rectangular profiled surface the main propagation mechanism was different from the other two surfaces.
Rather than finding a single streamer that propagates along the surface, the main propagation mechanism was due to re-ignition of new streamers in the neighboring corrugations.

In summary, we have presented an experimental and theoretical analysis of positive streamer discharges over various types of dielectric surface profiles.
Our analysis uses advanced fluid simulations together with experiments, and gives a qualitatively and quantitatively consistent picture of the nature of such discharges.
The presented insights on streamer-dielectric interaction over profiled surfaces opens up the possibility to tailor dielectric surfaces for various technological applications involving discharges over solid interfaces.

\section*{Acknowledgments}
This work is part of the project "Dielectric solutions for solid insulating components in eco-efficient medium-voltage switchgear" (project number: 321449) funded by the Research Council of Norway and ABB Electrification Norway AS.
The computations were performed on resources provided by UNINETT Sigma2 - the National Infrastructure for High Performance Computing and Data Storage in Norway.
The authors would like to thank Henriette Bilsbak for help with the experiments.
Data is available upon request.

\printbibliography

\end{document}